\documentclass[amsmath, aps, pra, twocolumn, floatfix]{revtex4-2}
\usepackage{graphicx}

\begin{document} 
	
\title{Optical trap for an atom around the midpoint between two coupled identical parallel optical nanofibers}
	
\author{Fam Le Kien}
\affiliation{Okinawa Institute of Science and Technology Graduate University, Onna, Okinawa 904-0495, Japan}
		
\author{S\'{i}le Nic Chormaic}
\affiliation{Okinawa Institute of Science and Technology Graduate University, Onna, Okinawa 904-0495, Japan}
	
\author{Thomas Busch}
\affiliation{Okinawa Institute of Science and Technology Graduate University, Onna, Okinawa 904-0495, Japan}
	
\date{\today}
	
\begin{abstract}
		
We study the trapping of a ground-state cesium atom in a small region around the midpoint between two coupled identical parallel optical nanofibers. We suggest to use a blue-detuned guided light field in the odd $\mathcal{E}_z$-sine array mode to produce an optical potential with a local minimum of exact zero at the midpoint between the nanofibers. 
We find that the effects of the van der Waals potential on the total trapping potential around the minimum point are not significant when the fiber separation distance and the power of the guided light field are large. For appropriate realistic parameters, a net trapping potential with a significant depth of about 1 mK, a large coherence time of several seconds, and a large recoil-heating-limited trap lifetime of several hours can be obtained. We investigate the dependencies of the trapping potential on the power of the guided light field, the fiber radius, the wavelength of light, and the fiber separation distance.

\end{abstract}
	
%\pacs{}
\maketitle
	
\section{Introduction}

Optical nanofibers are tapered fibers with a subwavelength diameter and significantly differing core and cladding refractive indices \cite{TongNat03}. Optical nanofibers have been investigated for various applications in nonlinear optics, atomic physics, quantum optics, and nanophotonics \cite{TongNat03,review2016,review2017,Nayak2018}. In particular, nanofibers have been used for trapping and optically interfacing cold atoms with guided light fields \cite{Dowling96,twocolor,Lacroute2012,Vetsch2010,Goban2012,Polzik2014,Aoki2015,Rolston2015,Laurat2016,Lacroute2012,onecolor,Sague08,Phelan13,LeKien09c,Reitz12,Schneeweiss2014,Daly2014}. 

A successful technique for trapping atoms near a nanofiber is to combine optical dipole forces of a blue- and a red-detuned guided light field \cite{Dowling96,twocolor}. This two-color trap scheme has been experimentally realized for laser-cooled alkali-metal atoms at about 200~nm above the nanofiber surface~\cite{Vetsch2010,Goban2012,Polzik2014,Aoki2015,Rolston2015,Laurat2016}. Other nanofiber-based atom traps have been proposed and investigated, such as a combination of the attractive potential of a red-detuned guided field and the repulsive potential of the centrifugal force~\cite{onecolor}, interference of higher-order modes~\cite{Sague08,Phelan13}, a diffracted laser field impinging perpendicularly to the fiber~\cite{LeKien09c}, a helical two-color trapping potential~\cite{Reitz12}, a combination of fictitious and real magnetic fields \cite{Schneeweiss2014}, and a nanofiber with a slot \cite{Daly2014,Daly2016}. 

Coupled waveguides play an important role in numerous optical devices such as multicore fibers, optical directional couplers, polarization splitters, ring resonators, and interferometers \cite{Snyder1983,Marcuse1989, Okamoto2006}. Recently, optical devices comprising of twisted or knotted nanofibers have been fabricated \cite{Glorieux2019}. 
Coupling between two nanofibers has been studied \cite{Glorieux2019,CMT} in the framework of the coupled mode theory \cite{Snyder1983,Marcuse1989,Okamoto2006}. It has been shown that the guided normal (array) modes of two coupled dielectric rods can be calculated by using the circular harmonics
expansion \cite{Wijngaard1973}. This method has been extended to multicore fibers \cite{Yamashita1985,Kishi1989,Huang1990,Chang1997a}. The propagation constant and the flux density of the field in a guided normal mode have been studied \cite{Wijngaard1973,Chang1997a,Huang1989}. The polarization patterns \cite{Chang1997a} and the mode cutoffs \cite{Chang1997b} have been investigated. In optomechanics, forces arising from internal illumination by light traveling in coupled waveguides have been studied \cite{Povinelli2005}, and light-guiding arrays of mechanically compliant glass nanospikes have been fabricated \cite{Wang2019}.

Recently, the spatial distributions of the fields in the guided array modes of two coupled parallel nanofibers have been examined \cite{tfexact}. It has been shown that the distribution of the field intensity in the odd $\mathcal{E}_z$-sine array mode has a local minimum of exact zero at the midpoint between the nanofibers \cite{tfexact}. This feature can be used to trap atoms with a single blue-detuned light field. In order to realize an optical trap for atoms in a small region around the midpoint between the nanofibers, we need to produce an optical potential that can dominate the surface-induced van der Waals potential in the trapping region. The parameters for the system must be realistic, while the characteristics of the obtained trapping potential must be appropriate for applications. The possibility of trapping atoms between two parallel nanofibers may open up new applications in nonlinear optics, quantum optics, and quantum information. 
While a setup with infinite, parallel nanofibers might be currently hard to realise, the configuration is also somewhat similar to a slotted fiber, as proposed for atom trapping in \cite{Daly2014} and demonstrated for nanoparticle trapping in \cite{Daly2016}. Therefore, it is desirable to study this issue in detail.

In this work, we study the possibility to trap a ground-state cesium atom in a small region around the midpoint between two coupled identical parallel optical nanofibers. We investigate the dependencies of the trapping potential on the power of the guided light field, the fiber radius, the wavelength of light, and the fiber separation distance. We show that a blue-detuned far-off-resonance field in the odd $\mathcal{E}_z$-sine array mode can produce a trapping potential around the midpoint between the nanofibers with a significant depth, a large coherence time, and a large trap lifetime.

The paper is organized as follows. In Sec. \ref{sec:model} we describe the model of two coupled identical parallel optical nanofibers and discuss the properties of the spatial field distribution in the odd $\mathcal{E}_z$-sine array mode. In Sec. \ref{sec:trap}, we calculate the trapping potential of an atom outside the nanofibers. Finally, we conclude in Sec. \ref{sec:summary}.

\section{Odd $\mathcal{E}_z$-sine array mode of two identical parallel nanofibers}
\label{sec:model}

%%%%%%%%%%%%%%%%%%%%%%% Figure 1
\begin{figure}[tbh]
	\begin{center}
		\includegraphics{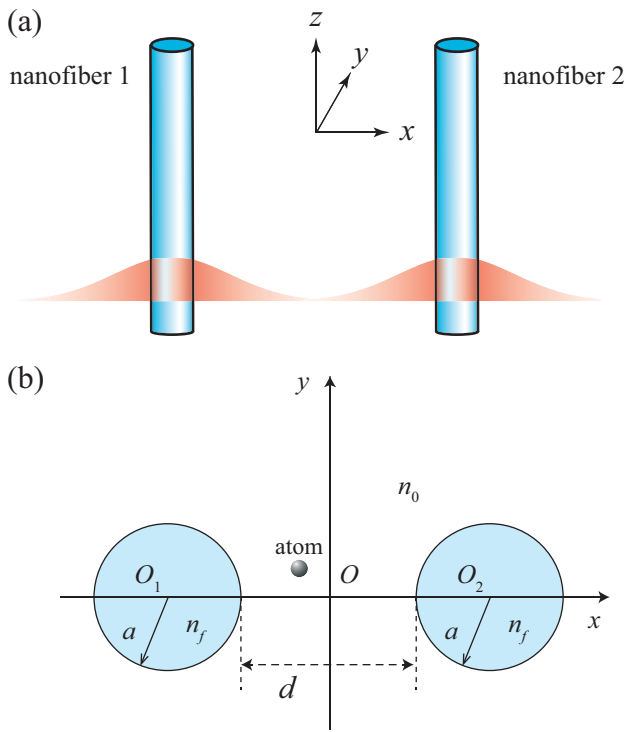}
	\end{center}		
	\caption{(Color online) Two coupled identical parallel optical nanofibers (a) and the geometry of the system (b). 	 
	}
	\label{fig1}
\end{figure}

We study two identical vacuum-clad, optical nanofibers that are aligned parallel to each other in the direction of the fiber axis $z$  (see  Fig.~\ref{fig1}). The fibers are labeled by the indices $j=1,2$. Each nanofiber $j$ can be treated as a dielectric cylinder with a radius $a_j=a$ and a refractive index $n_j=n_f>1$, surrounded by an infinite background of vacuum or air  with a refractive index $n_0=1$. The nanofiber diameters are a few hundreds of nanometers. 
Depending on the fiber size parameter $V=ka\sqrt{n_f^2-n_0^2}$, an individual nanofiber $j$ can support either a single or multiple modes. Here, $k=\omega/c$ is the wave number of light with optical frequency $\omega$ in free space. We are interested in the normal modes of the two-fiber system. We assume that the fibers are fixed in space and are, therefore, not interested in the van der Waals interaction between them. 

We introduce the global Cartesian coordinate system $\{x,y,z\}$. Here, the  $z$ axis is parallel to the  $z_1$ and $z_2$ axes of the fibers, the $x$ axis is perpendicular to the $z$ axis and connects the centers $O_1$ and $O_2$ of the fibers, and the  $y$ axis is perpendicular to the  $x$ and $z$ axes (see Fig.~\ref{fig1}). The plane $xy$ is the transverse (cross-sectional) plane of the fibers. The $x$ and $y$ axes are called the radial and tangential axes, respectively, of the two-fiber system [see Fig.~\ref{fig1}(b)]. The positions of the fiber centers $O_1$ and $O_2$ on the $x$ axis are $O_1=-(a+d/2)$ and $O_2=a+d/2$, where $d$ is the fiber separation distance. 
We introduce the polar coordinate system  $\{r,\varphi\}$ associated with the central Cartesian coordinate system $\{x,y\}$. For each individual fiber $j$, we introduce the local polar coordinate system $\{r_j,\varphi_j\}$.

We consider a light field with an optical frequency $\omega$ which propagates in the $+z$ direction with a propagation constant $\beta$. The electric and magnetic components of the field can be written as $\mathbf{E}=[\boldsymbol{\mathcal{E}}e^{-i (\omega t-\beta z)}+\mathrm{c.c.}]/2$ and $\mathbf{H}=[\boldsymbol{\mathcal{H}}e^{-i (\omega t-\beta z) t}+\mathrm{c.c.}]/2$, respectively, where $\boldsymbol{\mathcal{E}}$ and $\boldsymbol{\mathcal{H}}$ are the slowly varying complex envelopes. 

The normal modes  of the coupled fibers are called array modes. The exact theory for the guided normal modes of two parallel dielectric cylinders has been formulated in \cite{Wijngaard1973}. 
According to \cite{Wijngaard1973}, there are four kinds of normal modes, denoted  
as even $\mathcal{E}_z$-cosine, odd $\mathcal{E}_z$-cosine, even $\mathcal{E}_z$-sine, and odd $\mathcal{E}_z$-sine array modes. We are interested in the case where the fiber radii are small enough that no more than one normal mode of each of the four kinds can be supported. It has been shown that, for an odd $\mathcal{E}_z$-sine array mode of two coupled identical parallel fibers, the electric intensity distribution attains a local minimum of exactly zero at the two-fiber center (the midpoint between the nanofibers) \cite{tfexact}. 

We employ the theory of \cite{Wijngaard1973} to calculate the propagation constant and the spatial distribution of the field in an odd $\mathcal{E}_z$-sine array mode of two identical parallel vacuum-clad silica nanofibers \cite{tfexact}.
The key results of \cite{Wijngaard1973} for this mode are summarized in Appendix \ref{appendix}.
We solve equations (\ref{e6}) and use the expressions (\ref{e1}), (\ref{e2}), and (\ref{e8}) to calculate the components of the field. In our numerical calculations, the infinite number of circular harmonics is truncated
at a finite number $N_{\mathrm{max}}$ in the range from 9 to 19. The value of $N_{\mathrm{max}}$ is chosen such that the propagation constant converges and the boundary conditions are satisfied with a reasonable accuracy \cite{Wijngaard1973}. 
To calculate the refractive index $n_f$ of the silica nanofibers, we use the four-term Sellmeier formula for fused silica \cite{Malitson,Ghosh}. 

%%%%%%%%%%%%%%%%%%%%%%% Figure 2
\begin{figure}[tbh]
	\begin{center}
		\includegraphics{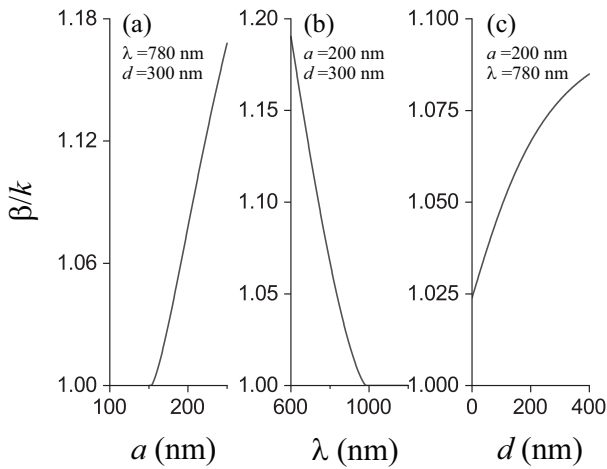}
	\end{center}
\caption{(Color online) Propagation constant $\beta$ of the odd $\mathcal{E}_z$-sine array mode, normalized to the free-space wave number $k$, as functions of (a) the fiber radius $a$, (b) the light field wavelength $\lambda$, and (c) the fiber separation distance $d$. The parameters are (a) $\lambda=780$ nm and  $d=300$ nm, (b) $a=200$ nm and $d=300$ nm, and (c) $a=200$ nm and $\lambda=780$ nm.	 
The refractive index of the fibers is $n_f=1.4537$ in (a) and (c), and is calculated from
the four-term Sellmeier formula for fused silica \cite{Malitson,Ghosh} in (b). 
The refractive index of the surrounding medium is $n_0=1$. 
	}
	\label{fig2}
\end{figure}

We plot in Fig.~\ref{fig2} the propagation constant $\beta$ of the odd $\mathcal{E}_z$-sine array mode as functions of the fiber radius $a$, the light field wavelength $\lambda$, and the fiber separation distance $d$. 
We observe from Figs.~\ref{fig2}(a) and \ref{fig2}(b) that the odd $\mathcal{E}_z$-sine array mode has cutoffs in the
dependencies of $\beta$ on $a$ and $\lambda$. The position of a cutoff is determined by the solution to the equation $\beta/k=1$ \cite{Wijngaard1973}.

%%%%%%%%%%%%%%%%%%%%%%% Figure 3
\begin{figure}[tbh]
	\begin{center}
		\includegraphics{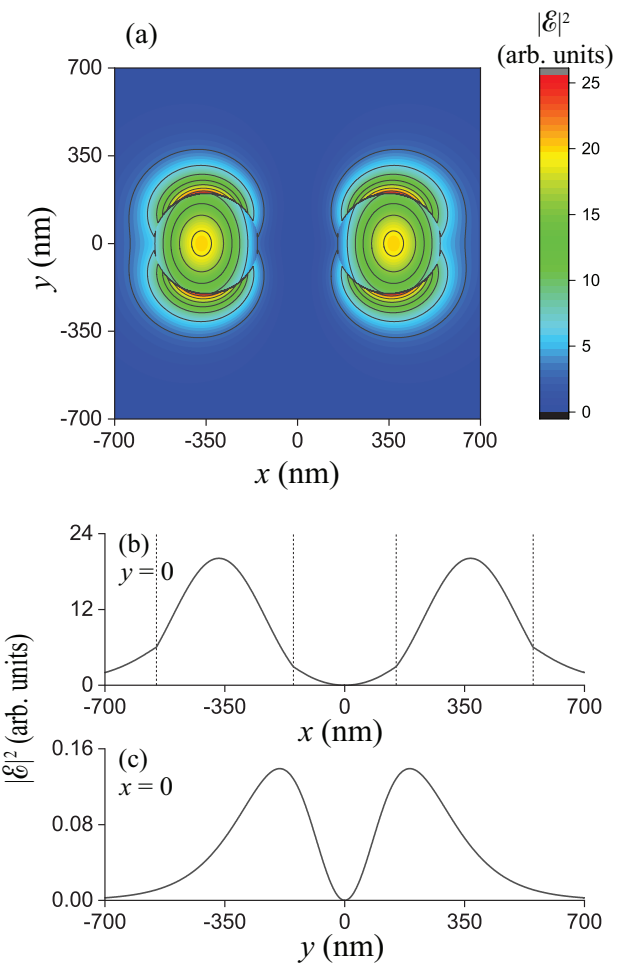}
	\end{center}
	\caption{(Color online) 
	    (a)	Cross-sectional profile and (b) $x$ and (c) $y$ axial profiles of the electric intensity distribution 
		$|\boldsymbol{\mathcal{E}}|^2$ of the field in the odd $\mathcal{E}_z$-sine array mode of two identical parallel nanofibers. The vertical dashed lines in (b) indicate the positions of the fiber surfaces on the $x$ axis. The wavelength of light is $\lambda=780$ nm, 	
		the fiber radius is $a=200$ nm, and the separation distance between the fibers is $d=300$ nm.  Other parameters are as in Fig.~\ref{fig2}.
	}
	\label{fig3}
\end{figure}

We display in Fig.~\ref{fig3} the cross-sectional profile and the axial profiles of the electric intensity distribution $|\boldsymbol{\mathcal{E}}|^2$ of the field in the odd $\mathcal{E}_z$-sine array mode of two identical parallel nanofibers. We see from the figure that $|\boldsymbol{\mathcal{E}}|^2$ is symmetric with respect to the principal axes $x$ and $y$. Figures \ref{fig3}(b) and \ref{fig3}(c) show that the electric field $\boldsymbol{\mathcal{E}}$ of the odd $\mathcal{E}_z$-sine array mode is exactly equal to zero at the midpoint $(x,y)=(0,0)$ between the nanofibers. This  feature of the odd $\mathcal{E}_z$-sine array mode can be used to trap ground-state atoms with a blue-detuned light field \cite{Nobel prizers a,Nobel prizers b,Nobel prizers c}. The existence of a local minimum of exact zero at the two-fiber center is due to the destructive interference between the fields of the individual fibers in the odd $\mathcal{E}_z$-sine array mode, and occurs for any fiber separation distance $d$.

\section{Trapping potential of an atom outside the nanofibers}
\label{sec:trap}

We consider an alkali-metal atom moving in the optical potential generated by an off-resonant guided light field in the odd $\mathcal{E}_z$-sine array mode outside the two nanofibers.

\subsection{Optical potential}

We assume that the atom is in the ground state and  the field is off resonance with the atom.
The optical potential $U_{\mathrm{opt}}$ of the atom in the field  is then given by \cite{Jackson} 
\begin{equation}\label{a1}
	U_{\mathrm{opt}}=-\frac{1}{4}\alpha |\boldsymbol{\mathcal{E}}|^2,
\end{equation}
where $\alpha=\alpha(\omega)$ is the real part of the scalar dynamical polarizability of the atom at the optical frequency $\omega$. 
The factor $1/4$ in Eq.~(\ref{a1}) results from the fact that the dipole of the atom is not a permanent dipole but is  induced by the field, giving $1/2$, and from the fact that the intensity is averaged over optical oscillations, giving another $1/2$. 

The function $\alpha(\omega)$ for a ground-state alkali-metal atom is given by \cite{Jackson}
\begin{equation}\label{a2}
	\alpha(\omega)=2\pi\epsilon_0 c^3\sum_j \frac{g_j}{g_a}
	\frac{A_{ja} (1-\omega^2/\omega_{ja}^2)}{(\omega_{ja}^2-\omega^2)^2+\gamma_{j}^2\omega^2}.
\end{equation}
Here, $g_j$ and $g_a$ are the statistical weights of the excited level $|j\rangle$ and the ground-state level $|a\rangle$, respectively,
$\omega_{ja}$ and $A_{ja}$ are the frequency and emission transition probability, respectively, of the spectral line $ja$, and $\gamma_{j}$ is the lifetime of the excited level $|j\rangle$.
We note that the vector polarizability is neglected in Eqs.~(\ref{a1}) and (\ref{a2}) and 
the tensor polarizability is vanishing for the ground-state alkali-metal atom \cite{Stark1,Stark2}.

To be specific, we consider atomic cesium.
A ground-state cesium atom has  two strong transitions, at 852 nm ($D_2$ line) and 894 nm ($D_1$ line). In order to trap the atom, we use the wavelength $\lambda=780$ nm, which is blue-detuned from the $D_1$ and $D_2$ lines. 
The dynamical polarizability of the ground-state cesium atom has been calculated numerically \cite{twocolor,Stark1,Stark2}.
To make simple calculations for the dynamical polarizability, 
we follow \cite{twocolor} and take into account the four most dominant lines of the atom, namely,
$\lambda_{1a}=852.113$ nm, $\lambda_{2a}=894.347$ nm, $\lambda_{3a}=455.528$ nm, 
and $\lambda_{4a}=459.317$ nm (see \cite{Cs}). The emission transition probabilities of these four lines are
$A_{1a}=3.276\times 10^7$ s$^{-1}$, $A_{2a}=2.87\times 10^7$ s$^{-1}$, 
$A_{3a}=1.88\times 10^6$ s$^{-1}$, and $A_{4a}=8\times 10^5$ s$^{-1}$.
The statistical weights of the upper states are $g_1=4$, $g_2=2$, $g_3=4$, and $g_4=2$, and that of the ground state is $g_a=2$. For the wavelength $\lambda=780$ nm, the polarizability of the atom is estimated to be $\alpha\cong -1709$ a.u.,
which is  negative and hence gives the repulsive optical potential $U_{\mathrm{opt}}=(1/4)|\alpha||\boldsymbol{\mathcal{E}}|^2$.

%%%%%%%%%%%%%%%%%%%%%%% Figure 4
\begin{figure}[tbh]
	\begin{center}
		\includegraphics{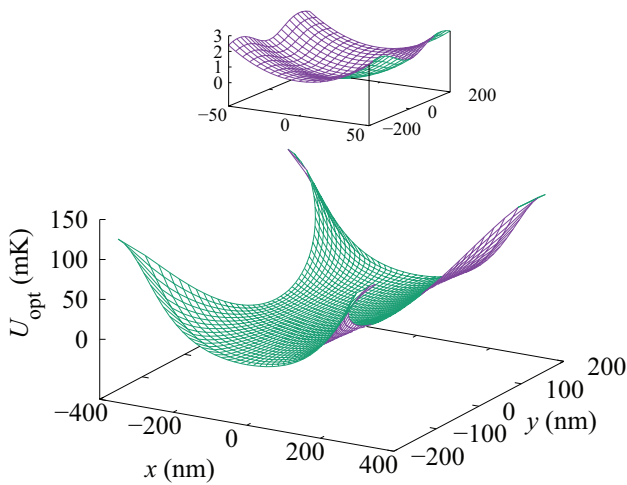}
	\end{center}
	\caption{(Color online) 
		Spatial profile of the optical potential $U_{\mathrm{opt}}$ of a ground-state cesium atom in the area $\{r_1,r_2>a; |x|<a+d/2; |y|< a  \}$, which lies between the two fibers. The inset shows $U_{\mathrm{opt}}$ in a small vicinity of the two-fiber center $O=(0,0)$. The trapping field is in the odd $\mathcal{E}_z$-sine array mode with the power of 100 mW. Other parameters are as in Fig.~\ref{fig3}.
	}
	\label{fig4}
\end{figure}

We plot in Fig.~\ref{fig4} the spatial profile of the optical potential $U_{\mathrm{opt}}$
of a ground-state cesium atom in the area $\{r_1,r_2>a; |x|<a+d/2; |y|< a  \}$, which lies between the two fibers. The inset of the figure shows $U_{\mathrm{opt}}$ in a small vicinity of the two-fiber center $O=(0,0)$. We observe that $U_{\mathrm{opt}}$ has a local minimum of exact zero at $O$. This feature can be used to trap atoms.
Note that, due to the geometry of the system, the optical potential is not cylindrically symmetric.

The rate of spontaneous scattering caused by a  light field $\boldsymbol{\mathcal{E}}$ is given by 
\begin{equation}\label{a3}
	\Gamma_{\mathrm{sc}}=\frac{1}{4\hbar}\kappa |\boldsymbol{\mathcal{E}}|^2,	
\end{equation}
where $\kappa=\kappa(\omega)$ is the imaginary part of the scalar dynamical polarizability of the atom.
The function $\kappa(\omega)$ for a ground-state alkali-metal atom is given by \cite{Jackson}  
\begin{equation}\label{a4}
	\kappa(\omega)
	=2\pi\epsilon_0 c^3\sum_j \frac{g_j}{g_a}
	\frac{A_{ja}\gamma_{ja}\omega/\omega_{ja}^2}{(\omega_{ja}^2-\omega^2)^2+\gamma_{j}^2\omega^2}.
\end{equation}

For atoms spending time in a motional quantum state $|\psi\rangle$, 
the average scattering rate is $\langle{\Gamma}_{\mathrm{sc}}\rangle\equiv\langle\psi|{\Gamma}_{\mathrm{sc}}|\psi\rangle=
(\kappa/{4\hbar})\langle|\boldsymbol{\mathcal{E}}|^2\rangle$ with $\langle|\boldsymbol{\mathcal{E}}|^2\rangle\equiv\langle\psi|(|\boldsymbol{\mathcal{E}}|^2)|\psi\rangle$.
The characteristic coherence time of the trap is
\begin{equation}\label{a5}
	\tau_{\mathrm{coh}}=\frac{1}{\langle{\Gamma}_{\mathrm{sc}}\rangle}. 
\end{equation}

A scattered photon imparts a recoil energy $E_{\mathrm{rec}}=(\hbar k)^2/2M$ to the atom, where $M$ is the mass of the atom. Therefore,
the absorption of incident photons and the emission of other photons result in a loss of atoms from the trapping potential. For a trap depth $U_D$, the trap lifetime due to recoil heating is given by  \cite{Julienne}
\begin{equation}\label{a6}
	\tau_{\mathrm{trap}}=\frac{U_D}{2E_{\mathrm{rec}}\langle\Gamma_{\mathrm{sc}}\rangle}.	
\end{equation}
When the light field frequency is far from the atomic resonances and the field at the trapping potential minimum is weak, the scattering rate is small and, therefore,
the coherence time and the trap lifetime are large. 

\subsection{van der Waals potential}

An atom near the surface of a medium undergoes a van der Waals force.  
The van der Waals potential of an atom at a radial position $r$ near the surface of a cylindrical dielectric rod of radius $a$, with $r>a$, is given by \cite{Boustimi}
\begin{eqnarray}\label{a7}
	V(r)&=&\frac{\hbar}{4\pi^3\epsilon_0}\sum_{n=-\infty}^{\infty}\int_0^{\infty} d\beta \,
	[\beta ^2K_n'^2(\beta r)\nonumber\\
	&&\mbox{}+(\beta ^2+n^2/r^2)K_n^2(\beta r)]\int_0^\infty d\xi\, \alpha(i\xi)G_n(i\xi,\beta ),
	\nonumber\\	
\end{eqnarray}
where
\begin{equation}\label{a8}
	G_n(i\xi,\beta )=\frac{[\epsilon(i\xi)-\epsilon_0]I_n(\beta a)I_n'(\beta a)}{\epsilon_0I_n(\beta a)K_n'(\beta a)-\epsilon(i\xi)I_n'(\beta a)K_n(\beta a)}.	
\end{equation}
Here $\epsilon(i\xi)$ is the dynamical dielectric function of the medium for the imaginary frequency $i\xi$, and $I_n$ and $K_n$ are the modified Bessel functions of the first and second kinds, respectively.
The van der Waals potential of a ground-state cesium atom near a silica fiber has been calculated numerically \cite{twocolor}.

For an atom near two identical parallel optical nanofibers, the van der Waals potential is given as
\begin{equation}\label{a9}
	U_{\mathrm{vdW}}=V(r_1)+V(r_2).
\end{equation}
Here, $r_1$  and $r_2$ are the distances from the atom to the fiber axes $z_1$ and $z_2$, respectively.

\subsection{Total potential}

The total potential of the atom is given as
\begin{equation}\label{a10}
	U=U_{\mathrm{opt}}+U_{\mathrm{vdW}}.
\end{equation} 

\begin{figure}[tbh]
	\begin{center}
		\includegraphics{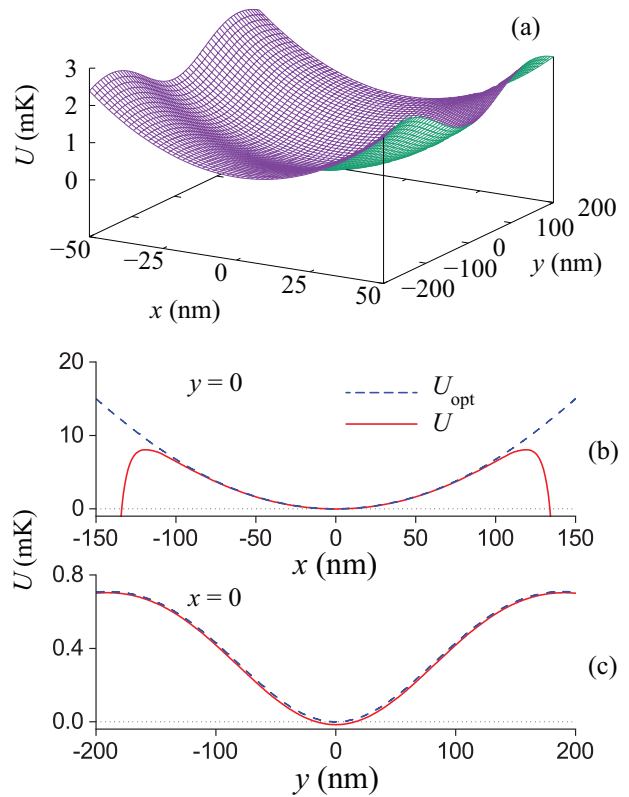}
	\end{center}
	\caption{(Color online) 
	(a)	Cross-sectional profile of the total potential $U$ of a ground-state cesium atom in a small area around the two-fiber center $O=(0,0)$.  	
	(b) and (c)	Axial profiles of the total potential $U$ (solid red lines) and the optical potential $U_{\mathrm{opt}}$ (dashes blue lines) of a ground-state cesium atom along the $x$ and $y$ axes.  	
		Parameters used are as in Figs.~\ref{fig3} and \ref{fig4}.
	}
	\label{fig5}
\end{figure}

We plot in Fig.~\ref{fig5}(a) the cross-sectional profile of the total potential $U$
of a ground-state cesium atom in the vicinity of the two-fiber center $O=(0,0)$.
In Figs.~\ref{fig5}(b) and \ref{fig5}(c), we depict the axial profiles of the total potential $U$ in the region between the nanofibers.
We observe from  the figures that $U(x,y)$ has a negative local minimum $U_{\mathrm{min}}\equiv U(0,0)=U_{\mathrm{vdW}}(0,0)<0$ at $O$. Comparison between the solid red and dashed blue lines of Figs.~\ref{fig5}(b) and \ref{fig5}(c) shows that the total potential $U$ is mainly determined by the optical potential $U_{\mathrm{opt}}$ in the vicinity of the two-fiber center $O$, which is positioned at a distance of 150 nm from the fiber surfaces in the case of the figures. 
This feature is a consequence of the fact that the effects of the van der Waals potential on the total trapping potential  around the minimum point are not significant when the fiber separation distance and the power of the guided light field are large enough. Like the optical potential $U_{\mathrm{opt}}$,  the total potential $U$ is not cylindrically symmetric due to the geometry of the system.
Figures \ref{fig5}(b) and \ref{fig5}(c) and the scales of their vertical axes show that
the depth of the profile of the potential along the $x$ axis is larger than that of the profile of the  potential along the $y$ axis. Therefore, the depth of the potential profile along the axis $y$ is the effective depth of the trapping potential.

%%%%%%%%%%%%%%%%%%%%%%% Figure 5

%%%%%%%%%%%%%%%%%%%%%%% Figure 6
\begin{figure}[tbh]
	\begin{center}
		\includegraphics{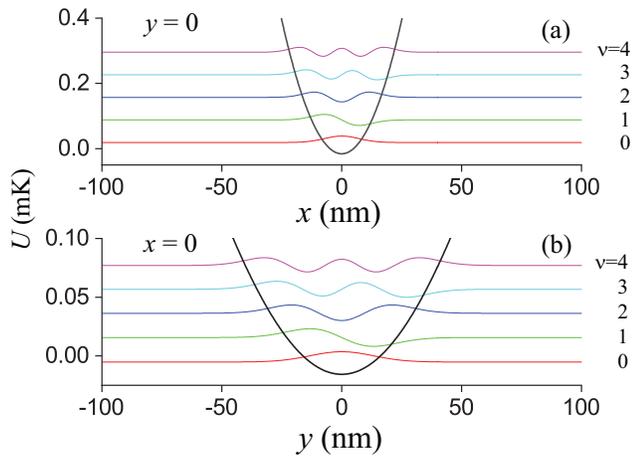}
	\end{center}
	\caption{(Color online)	
		Bound states for the first five levels ($\nu=0$, 1, 2, 3, and 4) of the one-dimensional motion of a ground-state cesium atom in the potential $U$ along the $x$ and $y$ axes.  	
		Parameters used are as in Figs.~\ref{fig3}--\ref{fig5}.
	}
	\label{fig6}
\end{figure}

The one-dimensional motion of the atom along the $x$ or $y$ axis can be treated as the motion of a particle in the  potential $U_x(x)=U(x,0)$ or $U_y(y)=U(0,y)$. In Fig.~\ref{fig6}, we plot the wave functions of the first five levels of the one-dimensional motion of the atom in the  potentials $U_x$ and $U_y$ for the case of Fig.~\ref{fig5}. We find that the trapping frequencies are $\omega_x/2\pi\cong 1441$ kHz and $\omega_y/2\pi\cong 438$ kHz, comparable to the characteristic values for the single-nanofiber two-color traps \cite{Vetsch2010,Goban2012,Polzik2014,Aoki2015,Rolston2015,Laurat2016}. 
The spacing between the  the energies of the ground state and the first excited state is  roughly equal to $\hbar\omega_x\cong k_B\times 69$ $\mu$K for the motion along the $x$ axis and $\hbar\omega_y\cong k_B\times 21$ $\mu$K for the motion along the $y$ axis. The characteristic sizes of the corresponding motional ground states are $\Delta x=\sqrt{\hbar/2M\omega_x}\cong 5.1$ nm and $\Delta y=\sqrt{\hbar/2M\omega_y}\cong 9.3$ nm.

The scattering rates for the ground states of the atomic motions along the $x$ and $y$ axes in the case of Fig.~\ref{fig5} are found to be $\langle\Gamma_{\mathrm{sc}}\rangle_x \cong 0.17$ s$^{-1}$ and $\langle\Gamma_{\mathrm{sc}}\rangle_y \cong 0.05$ s$^{-1}$. 
The scattering rate $\langle\Gamma_{\mathrm{sc}}\rangle$ of the atom in the trap is estimated as the maximal value of  $\langle\Gamma_{\mathrm{sc}}\rangle_x$ and $\langle\Gamma_{\mathrm{sc}}\rangle_y$, and is, in the case of Fig.~\ref{fig5}, given as $\langle\Gamma_{\mathrm{sc}}\rangle\cong 0.17$ s$^{-1}$. 
The corresponding coherence time is $\tau_{\mathrm{coh}}\equiv 1/\langle\Gamma_{\mathrm{sc}}\rangle \cong 5.8$ s. The trap depth is estimated as the minimal value of the trap depths for the motions along the $x$ and $y$ axes, and is, in the case of Fig.~\ref{fig5}(c), equal to $U_D\cong 0.7$ mK, comparable to the depths of the single-nanofiber two-color traps  \cite{Vetsch2010,Goban2012,Polzik2014,Aoki2015,Rolston2015,Laurat2016}. The corresponding recoil-limited trap lifetime is $\tau_{\mathrm{trap}}\cong 4.8$ h. This value is much larger than the recoil-limited lifetimes of 100 s and 30 s estimated for the single-nanofiber two-color traps in the experiments \cite{Vetsch2010} and \cite{Goban2012}, respectively.

\subsection{Dependencies of the trapping potential on the parameters of the fibers and the light field}

The trapping potential $U$ depends on the light field power $P$, the fiber radius $a$, the light field wavelength $\lambda$, and the fiber separation distance $d$.

%%%%%%%%%%%%%%%%%%%%%%% Figure 7
\begin{figure}[tbh]
	\begin{center}
		\includegraphics{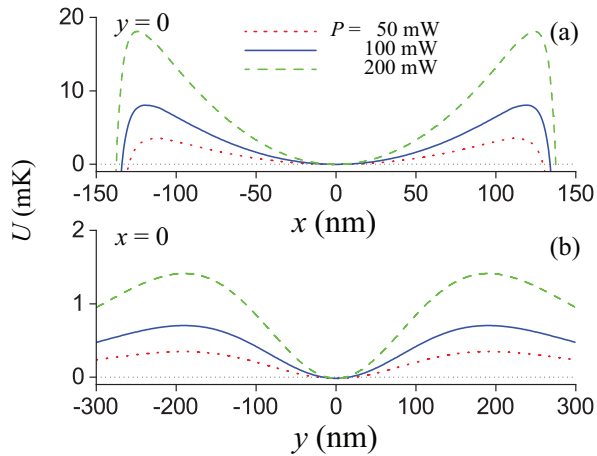}
	\end{center}
	\caption{(Color online) 
		Axial profiles of the total potential $U$ of a ground-state cesium atom for different values $P=50$ mW (dotted red lines), 100 mW (solid blue lines), and 200 mW (dashed green lines) of the light field power. The fiber radius is $a=200$ nm, the wavelength of light is $\lambda=780$ nm, and the fiber separation distance is $d=300$ nm. Other parameters are as in Figs.~\ref{fig3}--\ref{fig5}.
	}
	\label{fig7}
\end{figure}

The power of the guided light is given by $P=(1/2)\int \mathrm{Re}\,[\boldsymbol{\mathcal{E}}\times\boldsymbol{\mathcal{H}}^*]_z d\mathbf{r}$, where $\mathbf{r}=(x,y)$
and $\int d\mathbf{r}=\int_{-\infty}^{\infty}dx\int_{-\infty}^{\infty}dy$.
We plot in Fig.~\ref{fig7} the trapping potential $U$ for different values of the power $P$ of the guided light.
It is clear from the figure that the magnitude and the depth of the trapping potential are almost linearly proportional to the power. This feature is a result of the fact that, when the position $\mathbf{r}$ is not too close to the fiber surfaces, the total potential $U(\mathbf{r})$ is mainly determined by the optical potential $U_{\mathrm{opt}}(\mathbf{r})$, which is linearly proportional to the power of light [see Eq.~(\ref{a1})].

%%%%%%%%%%%%%%%%%%%%%%% Figure 8
\begin{figure}[tbh]
	\begin{center}
		\includegraphics{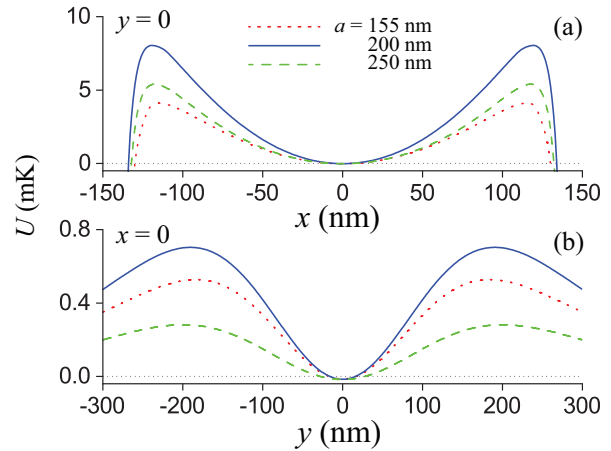}
	\end{center}
	\caption{(Color online) 
		Axial profiles of the total potential $U$ of a ground-state cesium atom for different values $a=155$ nm (dotted red lines), 200 nm (solid blue lines), and 250 nm (dashed green lines) of the fiber radius. The wavelength of light is $\lambda=780$ nm, the separation distance between the fibers is $d=300$ nm, and the power of light is $P=100$ mW. Other parameters are as in Figs.~\ref{fig3}--\ref{fig5}.
	}
	\label{fig8}
\end{figure}

We plot in Fig.~\ref{fig8} the trapping potential $U$ for different values of the fiber radius $a$.
We observe from the figure that, when $a$ is small (see the dotted red lines) or large (see the dashed green lines), $U$ is shallow. These features are consequences of the wide spread of the guided field outside a thin fiber  and the tight confinement of the guided field inside a thick fiber.

%%%%%%%%%%%%%%%%%%%%%%% Figure 9
\begin{figure}[tbh]
	\begin{center}
		\includegraphics{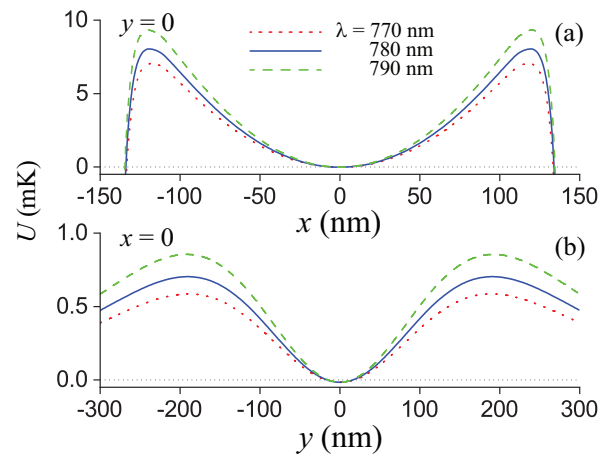}
	\end{center}
	\caption{(Color online) 
		Axial profiles of the total potential $U$ of a ground-state cesium atom for different values $\lambda=770$ nm (dotted red lines), 780 nm (solid blue lines), and 790 nm (dashed green lines) of the wavelength of light.  	
		The fiber radius is $a=200$ nm, the separation distance between the fibers is $d=300$ nm, and the power of light is $P=100$ mW. Other parameters are as in Figs.~\ref{fig3}--\ref{fig5}.
	}
	\label{fig9}
\end{figure}

We plot in Fig.~\ref{fig9} the trapping potential $U$ for different values of the wavelength of light $\lambda$.
The figure shows that, when $\lambda$ is closer to  the resonance with the atom (852 nm for the atomic cesium $D_2$ line), $U$ is deeper. These features appear as the consequences of the dependence of the atomic polarizability on the wavelength of a detuned light field. Note that the dependence of the trapping potential on the wavelength of light occurs through not only the atomic polarizability but also 
the mode profile and, consequently, the field intensity distribution.

%%%%%%%%%%%%%%%%%%%%%%% Figure 10
\begin{figure}[tbh]
	\begin{center}
		\includegraphics{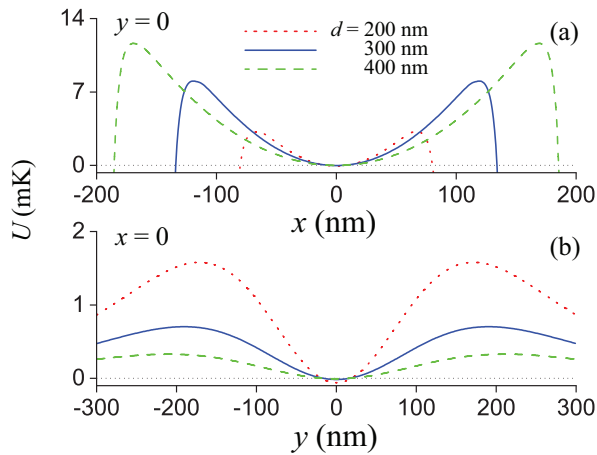}
	\end{center}
	\caption{(Color online) 
		Axial profiles of the total potential $U$ of a ground-state cesium atom for different values $d=200$ nm (dotted red lines), 300 nm (solid blue lines), and 400 nm (dashed green lines) of the separation distance between the fibers. The fiber radius is $a=200$ nm, the wavelength of light is $\lambda=780$ nm, and the power of light is $P=100$ mW. Other parameters are as in Figs.~\ref{fig3}--\ref{fig5}.
	}
	\label{fig10}
\end{figure}

We plot in Fig.~\ref{fig10} the trapping potential $U$ for different values of the fiber separation distance $d$.
The figure shows that an increase in the separation distance $d$ leads to an increase in the depth of the trap along the $x$ direction [see Fig.~\ref{fig10}(a)] and to a decrease in the depth of the trap along the $y$ direction [see Fig.~\ref{fig10}(b)]. We also observe that an increase in $d$ leads to an almost linear increase in the width of the trapping potential along the $x$ axis but does not affect much the width of the trapping potential along the $y$ axis.

\section{Summary}
\label{sec:summary}

In this work, we have studied the trapping potential of a ground-state cesium atom in a small region around the midpoint between two coupled identical parallel optical nanofibers. We have suggested to use a blue-detuned guided light field in the odd $\mathcal{E}_z$-sine array mode of the two nanofibers to produce an optical potential with a local minimum at the midpoint between the  nanofibers. 
The vanishing of the field at the two-fiber center is a result of the destructive interference between the coupled modes of the individual nanofibers. We have demonstrated that the effects of the van der Waals potential on the total trapping potential  around the minimum point are not significant when the fiber separation distance and the power of the guided light field are large. We have shown that, for appropriate realistic parameters, a net potential with a significant depth, a large coherence time, and a large trap lifetime can be obtained. We have demonstrated, for example, that a pair of  200-nm-radius silica fibers carrying 100 mW of 780-nm-wavelength  light in the odd $\mathcal{E}_z$-sine array mode gives for cesium atoms a trap depth of 0.7 mK, a coherence time of 5.8 s, and a recoil-heating-limited trap lifetime of 4.8 h. Due to the geometry of the system, the trapping potential is not cylindrically symmetric. The depth of the trapping potential profile along the tangential direction $y$ is smaller than that of the trapping potential profile along the radial direction $x$ and is, hence, the effective trap depth. The dependencies of the trapping potential on the power of the guided light field, the fiber radius, the wavelength of light, and the fiber separation distance have been investigated. It has been shown that, at a position that is not too close to the fiber surfaces, the total potential is mainly determined by the optical potential and, hence, is linearly proportional to the power of light. When the fiber radius is small  or large enough, the potential at the midpoint between the nanofibers is shallow due to the wide spread of the guided field outside a thin fiber and the tight confinement of the guided field inside a thick fiber. We have observed that the wavelength of light affects the trapping potential through the atomic polarizability and the field intensity distribution. We have found that the depth of the trapping potential decreases with increasing fiber separation distance.

\begin{acknowledgments}
This work was supported by the Okinawa Institute of Science and Technology Graduate University and by the Grant-in-Aid for Scientific Research (C) 20K03795.
\end{acknowledgments}

%%%%%%%%%%%%%%%%%%%%%%%%%%%%%%%%%%%%%%%%%%
%%%%%%%%%%%%%%%%%%%%%%%%%%%%%%%%%%%%%%%%%%

\appendix

\section{Odd $\mathcal{E}_z$-sine array mode}
\label{appendix}

According to \cite{Wijngaard1973}, for an odd $\mathcal{E}_z$-sine guided normal (array) mode, the longitudinal components $\mathcal{E}_z$ and $\mathcal{H}_z$ of the electric and magnetic parts, respectively, of the field  are given, inside fiber $j=1,2$, as
\begin{eqnarray}\label{e1}
	\mathcal{E}_z&=&\sum_{n=0}^\infty E_{nj}J_n(hr_j)\sin n\varphi_j,\nonumber\\
	\mathcal{H}_z&=&\sum_{n=0}^\infty F_{nj}J_n(hr_j)\cos n\varphi_j,
\end{eqnarray}
and, outside the two fibers, as 
\begin{eqnarray}\label{e2}
	\mathcal{E}_z&=&\sum_{j=1}^2\sum_{n=0}^\infty
	G_{nj}K_n(qr_j)\sin n\varphi_j,\nonumber\\
	\mathcal{H}_z&=&\sum_{j=1}^2\sum_{n=0}^\infty
	H_{nj}K_n(qr_j)\cos n\varphi_j.
\end{eqnarray}
Here, we have introduced the fiber parameters
\begin{equation}\label{e3}
	h=\sqrt{k^2n_f^2-\beta^2},\qquad
	q=\sqrt{\beta^2-k^2n_0^2},
\end{equation}
which determine the scales of the spatial variations of the field both inside and outside the fibers.
In Eqs.~(\ref{e1}) and (\ref{e2}), $E_{nj}$, $F_{nj}$, $G_{nj}$, and $H_{nj}$ are the mode expansion coefficients, $J_n$ represents the Bessel functions of the first kind, and $K_n$ represents the modified Bessel functions of the second kind.

The symmetry relations for the coefficients $E_{mj}$, $F_{mj}$, $G_{mj}$, and $H_{mj}$ with $j=1,2$ are \cite{Wijngaard1973}
\begin{eqnarray}\label{e4}
	E_{m2}&=& (-1)^m E_{m1}, \qquad F_{m2}=(-1)^m F_{m1},\nonumber\\
	G_{m2}&=& (-1)^m G_{m1}, \qquad H_{m2}=(-1)^m H_{m1}.
\end{eqnarray}  
The coefficients $E_{n1}$ and $F_{n1}$ are given by the equations \cite{Wijngaard1973}
\begin{eqnarray}\label{e5}
	J_n(u)E_{n1}&=&K_n(w)G_{n1}+I_n(w)\sum_{m=0}^\infty g_{nm}G_{m1},\nonumber\\
	J_n(u)F_{n1}&=&K_n(w)H_{n1}+I_n(w)\sum_{m=0}^\infty f_{nm}H_{m1}.\qquad
\end{eqnarray} 
The coefficients $G_{n1}$ and $H_{n1}$ are nonzero solutions of the equations \cite{Wijngaard1973}
\begin{eqnarray}\label{e6}
	&&	n\bigg(\frac{1}{u^2}+\frac{1}{w^2}\bigg)\bigg[ K_n(w)G_{n1}+I_n(w)\sum_{m=0}^\infty g_{nm}G_{m1}\bigg]
	\nonumber\\&&\mbox{}  
	-\frac{\omega\mu_0}{\beta}\bigg[\frac{J'_n(u)}{uJ_n(u)}
	+\frac{K'_n(w)}{wK_n(w)}\bigg] K_n(w)H_{n1}
	\nonumber\\&&\mbox{}	 
	-\frac{\omega\mu_0}{\beta }\bigg[\frac{J'_n(u)}{uJ_n(u)}
	+\frac{I'_n(w)}{wI_n(w)}\bigg]I_n(w)\sum_{m=0}^\infty f_{nm}H_{m1}
	=0,
	\nonumber\\
	&&  n\bigg(\frac{1}{u^2}+\frac{1}{w^2}\bigg)\bigg[K_n(w)H_{n1}
	+I_n(w)\sum_{m=0}^\infty f_{nm}H_{m1}\bigg]
	\nonumber\\&&\mbox{}
	-\frac{\omega\epsilon_0}{\beta}\bigg[\frac{n_1^2J'_n(u)}{uJ_n(u)}
	+\frac{n_0^2K'_n(w)}{wK_n(w)}\bigg]K_n(w)G_{n1}
	\nonumber\\&&\mbox{}
	-\frac{\omega\epsilon_0}{\beta}\bigg[\frac{n_1^2J'_n(u)}{uJ_n(u)}
	+\frac{n_0^2I'_n(w)}{wI_n(w)}\bigg]
	I_n(w)\sum_{m=0}^\infty g_{nm}G_{m1}
	=0.
	\nonumber\\
\end{eqnarray}
Here, we have introduced the coefficients
\begin{eqnarray}\label{e7}
	f_{nm}&=& K_{m+n}(qW)+ K_{m-n}(qW) \mbox{ for } n>0,\nonumber\\
	f_{0,m}&=& K_{m}(qW),\nonumber\\
	g_{nm}&=& -K_{m+n}(qW)+ K_{m-n}(qW),
\end{eqnarray}
where $W=d+a_1+a_2$ is the distance between the fiber centers.

In terms of the longitudinal components $\mathcal{E}_z$ and $\mathcal{H}_z$  of the field,
the transverse components $\mathcal{E}_{x,y}$ and $\mathcal{H}_{x,y}$ are given as \cite{Snyder1983,Marcuse1989,Okamoto2006}
\begin{eqnarray}\label{e8}
	\mathcal{E}_x&=&\frac{i\beta}{k^2n_{\mathrm{ref}}^2-\beta^2}\left(\frac{\partial}{\partial x}\mathcal{E}_z+\frac{\omega\mu_0}{\beta}\frac{\partial}{\partial y}\mathcal{H}_z\right),\nonumber\\
	\mathcal{E}_y&=&\frac{i\beta}{k^2n_{\mathrm{ref}}^2-\beta^2}\left(\frac{\partial}{\partial y}\mathcal{E}_z-\frac{\omega\mu_0}{\beta}\frac{\partial}{\partial x}\mathcal{H}_z\right),\nonumber\\
	\mathcal{H}_x&=&\frac{i\beta}{k^2n_{\mathrm{ref}}^2-\beta^2}\left(\frac{\partial}{\partial x}\mathcal{H}_z-\frac{\omega\epsilon_0n_{\mathrm{ref}}^2}{\beta}\frac{\partial}{\partial y}\mathcal{E}_z\right),\nonumber\\
	\mathcal{H}_y&=&\frac{i\beta}{k^2n_{\mathrm{ref}}^2-\beta^2}\left(\frac{\partial}{\partial y}\mathcal{H}_z+\frac{\omega\epsilon_0n_{\mathrm{ref}}^2}{\beta}\frac{\partial}{\partial x}\mathcal{E}_z\right).
\end{eqnarray}
Here, $n_{\mathrm{ref}}$ is the spatial distribution of the refractive index of the two-fiber system,
that is, $n_{\mathrm{ref}}=n_f$ inside the fibers and $n_{\mathrm{ref}}=n_0$ outside the two fibers.

The dispersion equation for the propagation constant of the mode is $\Delta=0$, where $\Delta$ is the determinant of the system of linear equations (\ref{e6}) for $G_{nj}$ and $H_{nj}$. The solution to the equation $\Delta=0$ determines the propagation constant $\beta$, which allows us to calculate the other fiber parameters $h$ and $q$ [see Eqs.~(\ref{e3})]. 

We make $E_{nj},F_{nj}$, $G_{nj}$, and $H_{nj}$ real-valued coefficients by omitting a common global phase.   
Then, for the electric part of the field,  we have   
$\mathcal{E}_z^*=\mathcal{E}_z$, $\mathcal{E}_x^*=-\mathcal{E}_x$, and $\mathcal{E}_y^*=-\mathcal{E}_y$.
Thus, the longitudinal component $\mathcal{E}_z$ of the field in a guided normal mode is $\pi/2$ out of phase with respect to the transverse components $\mathcal{E}_x$ and $\mathcal{E}_y$. This is a typical feature of transversely confined light fields \cite{Snyder1983,Marcuse1989,Okamoto2006,Lodahl2017}. 

It follows from the relations (\ref{e4}) and Eqs.~(\ref{e1}), (\ref{e2}), and (\ref{e8}) that the field components of the odd $\mathcal{E}_z$-sine array modes satisfy the relations
$\mathcal{E}_x(x,-y)=-\mathcal{E}_x(x,y)$, $\mathcal{E}_z(x,-y)=-\mathcal{E}_z(x,y)$,
$\mathcal{E}_y(x,y)=-\mathcal{E}_y(-x,y)$, and $\mathcal{E}_z(x,y)=-\mathcal{E}_z(-x,y)$, indicating the antisymmetry of $\mathcal{E}_x$ and $\mathcal{E}_z$ about the $x$ axis and that of $\mathcal{E}_y$ and $\mathcal{E}_z$ about the $y$ axis. It follows from these relations that, for the odd $\mathcal{E}_z$-sine array mode, the electric field at the two-fiber center $(x,y)=(0,0)$ is zero, that is, $\boldsymbol{\mathcal{E}}(0,0)=0$. This feature of the odd $\mathcal{E}_z$-sine array mode can be used to produce a local minimum of a blue-detuned optical dipole potential to trap ground-state atoms \cite{Nobel prizers a,Nobel prizers b,Nobel prizers c} or
a local minimum of a ponderomotive optical Rydberg-electron potential
to trap Rydberg atoms \cite{ponderomotive 1,ponderomotive 2}.

%%%%%%%%%%%%%%%%%%%%%%%%%%%%%%%%%%%%%%%%%%
%%%%%%%%%%%%%%%%%%%%%%%%%%%%%%%%%%%%%%%%%%

\end{document}